\documentclass[pra,aps,onecolumn,nopacs,superscriptaddress,nofootinbib]{revtex4}
\usepackage[version=3]{mhchem} 
\usepackage{enumitem}
\usepackage{verbatim}
\usepackage[T1]{fontenc}
\usepackage[latin1]{inputenc}
\usepackage{lmodern}
\usepackage{graphicx}
\usepackage{dcolumn}
\usepackage{bm}
\usepackage{amsmath}
\usepackage{amssymb}
\usepackage{pst-all}
\usepackage{psfrag}
\usepackage{epsfig}
\usepackage{color}
\usepackage{upgreek}

\begin{document}

\title{Analytical Modeling of Graphene Plasmons}

\author{Renwen~Yu}
\affiliation{ICFO-Institut de Ciencies Fotoniques, The Barcelona Institute of Science and Technology, 08860 Castelldefels (Barcelona), Spain}
\author{Joel~D.~Cox}
\affiliation{ICFO-Institut de Ciencies Fotoniques, The Barcelona Institute of Science and Technology, 08860 Castelldefels (Barcelona), Spain}
\author{J.~R.~M.~Saavedra}
\affiliation{ICFO-Institut de Ciencies Fotoniques, The Barcelona Institute of Science and Technology, 08860 Castelldefels (Barcelona), Spain}
\author{F.~Javier~Garc\'{\i}a~de~Abajo}
\affiliation{ICFO-Institut de Ciencies Fotoniques, The Barcelona Institute of Science and Technology, 08860 Castelldefels (Barcelona), Spain}
\affiliation{ICREA-Instituci\'o Catalana de Recerca i Estudis Avan\c{c}ats, Passeig Llu\'{\i}s Companys 23, 08010 Barcelona, Spain}
\email{javier.garciadeabajo@nanophotonics.es}

\date{\today}

\def\ii{{i}}  \def\ee{{e}} \def\Eb{{\bf E}}  \def\Bb{{\bf B}}  \def\Hb{{\bf H}}  \def\jb{{\bf j}} \def\Rb{{\bf R}}  \def\rb{{\bf r}}  \def\Qb{{\bf Q}}  \def\qb{{\bf q}} \def\vb{{\bf v}}  \def\pb{{\bf p}}  \def\kb{{\bf k}} \def\xx{\hat{\bf x}}  \def\yy{\hat{\bf y}}  \def\zz{\hat{\bf z}}  \def\kk{{\hat{\bf k}_\parallel}}
\def\th{{\vec{\theta}}}  \def\fE{\vec{\mathcal{E}}}  \def\Mb{{\bf M}}  \def\zetav{\vec{\zeta}}  \def\fEa{\mathcal{E}}
\def\vF{{v_{\rm F}}}  \def\kF{{k_{\rm F}}}  \def\EF{{E_{\rm F}}} \def\kpar{{k_\parallel}}  \def\kbpar{{{\bf k}_\parallel}}

\begin{abstract}
The two-dimensionality of graphene and other layered materials can be exploited to simplify the theoretical description of their plasmonic and polaritonic modes. We present an analytical theory that allows us to simulate these excitations in terms of plasmon wave functions (PWFs). Closed-form expressions are offered for their associated extinction spectra, involving only two real parameters for each plasmon mode and graphene morphology, which we calculate and tabulate once and for all. Classical and quantum-mechanical formulations of this PWF formalism are introduced, in excellent mutual agreement for armchaired islands with $>10\,$nm characteristic size. Examples of application are presented to predict both plasmon-induced transparency in interacting nanoribbons and excellent sensing capabilities through the response to the dielectric environment. We argue that the PWF formalism has general applicability and allows us to analytically describe a wide range of 2D polaritonic behavior, thus facilitating their use for the design of actual devices.
\end{abstract}
\maketitle

{\bf Keywords:} graphene plasmons, 2D polaritonics, electromagnetic modeling, plasmon wave function, plasmon-induced transparency, optical sensing


Plasmons are collective oscillations of conduction electrons found in different materials, where they interact strongly with light and can confine it down to nanoscale spatial regions to generate enormous optical field intensity enhancement \cite{LSB03}. These extraordinary properties are of paramount importance for a wide range of applications, such as optical sensing and modulation \cite{KWK97,NE97,XBK99,M05_2,paper274}, the enhancement of nonlinear optical processes \cite{DN07,PN08}, photocatalysis \cite{SLL12_2,GZT12,MLL13,MLS13,C14,MZG14}, and photothermal therapies \cite{QPA08,HSB03}. In these applications, precise spectral positioning of plasmon resonances is needed to achieve optimal performance. This is commonly achieved by fabricating noble metal nanostructures with specific sizes and morphologies. However, despite being the workhorse of plasmonics research, noble metals unfortunately present relatively large inelastic losses, thus limiting plasmon lifetimes in metallic nanostructures \cite{JC1972} and leading to a severe reduction in optical confinement. Additionally, the large number of electrons involved in the plasmons of metallic nanostructures limits the ways in which we can influence them in a dynamical fashion.

Recently, highly-doped graphene has emerged as an outstanding plasmonic material \cite{WSS06,HD07,JBS09,JGH11,FAB11,SKK11,paper196,FRA12,YLC12,YLL12,paper212,BJS13,YLZ13,YAL14} that simultaneously provides strong field confinement with relatively lower loss \cite{WLG15}. More importantly, plasmons in graphene are sustained by a small number of charge carriers compared to those of traditional noble metals, a property that makes them amenable to display new phenomena, including an unprecedented electro-optical response. Indeed, active tunability of the plasmon resonance frequency has been achieved via electrical gating \cite{JGH11,FAB11,SKK11,paper196,FRA12,YLC12,paper212,BJS13,YLZ13}. Additionally, many of the aforementioned applications that were first realized using noble metal plasmons have now been realized using a tunable graphene platform \cite{YAL14,WLG15,paper256,paper277}. However, the design of graphene-based plasmonic devices requires accurate modeling of their optical response, often necessitating time-consuming numerical simulations.

Here we present an analytical model based on the so-called plasmon wave functions (PWFs), which can accurately predict the optical response associated with plasmonic resonances sustained in doped graphene structures with arbitrary shape and size. Actually, the present model can describe plasmons in any two-dimensional (2D) structure using only two real-valued parameters, and thus constitutes a powerful tool that can be used in the design of graphene-based nanoplasmonic devices. We further compare the concept of the PWFs, which are the induced charge density profiles associated with confined plasmonic modes, with rigorous classical electromagnetic and atomistic quantum-mechanical (QM) models for nanostructured graphene. As a proof-of-concept, we demonstrate the use of our analytical model in the study of various applications for graphene-based nanostructures, namely, plasmon-induced transparency (PIT) and refractive index sensing.

\begin{figure}
\begin{centering}
\includegraphics[width=1\textwidth]{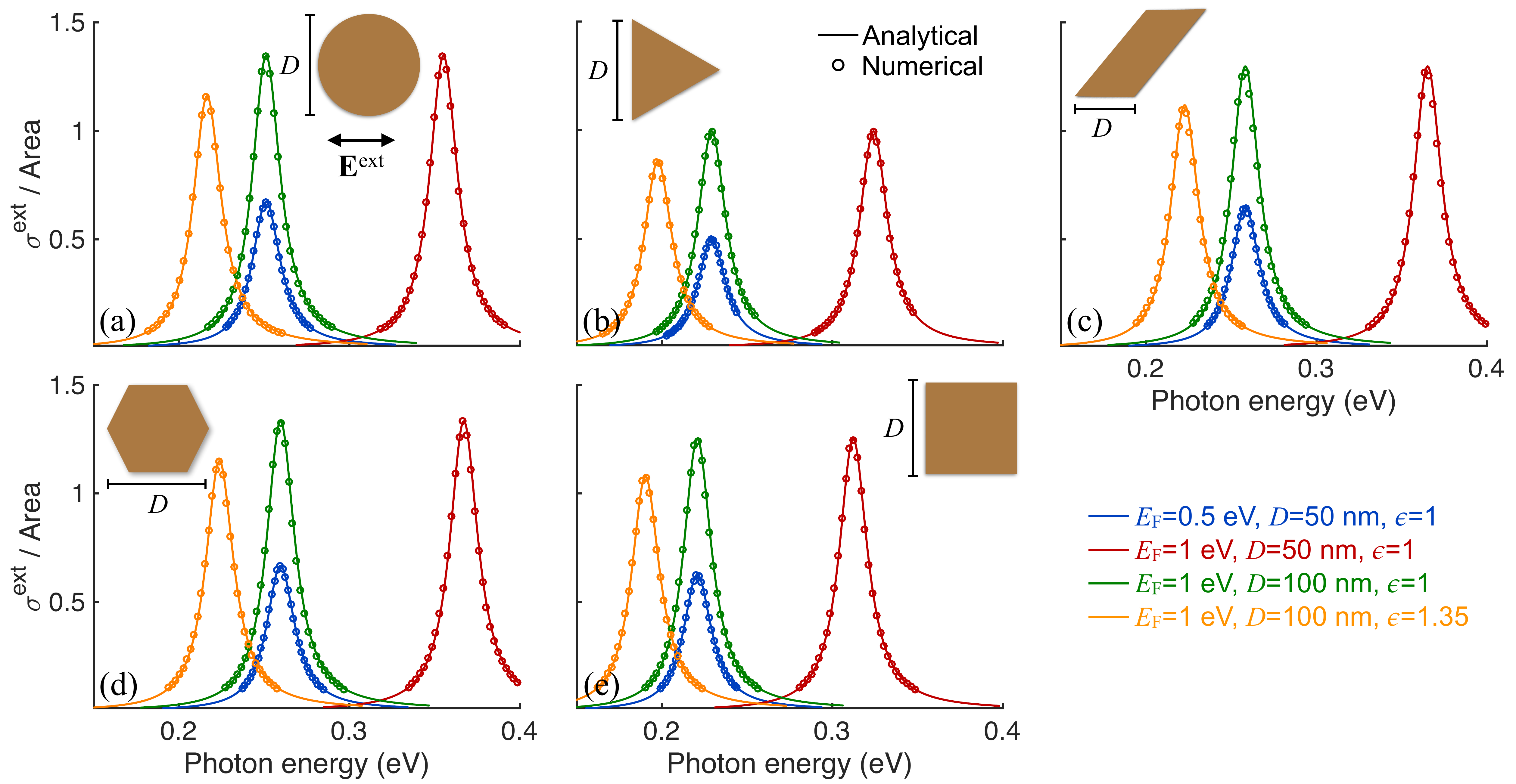}
\par\end{centering}
\caption{\textbf{Analytical description of plasmons in graphene islands of different morphology.} We present the extinction cross-section for (a) disks, (b) triangles, (c) ribbons, (d) hexagons, and (e) squares. In each figure, analytical results obtained using the plasmon wave function (PWF) of the lowest-order dipolar mode (solid curves) are compared with numerical simulations (symbols) for different combinations of characteristic size $D$, graphene Fermi energy $\EF$, and surrounding dielectric permittivity $\epsilon$ (see color-coded legend). In all calculations we describe the graphene surface conductivity using the Drude model (Eq.\ (\ref{Drude})), adopting a phenomenological inelastic damping energy $\hbar\tau^{-1}=20\,$meV (i.e., $\tau\approx33\,$fs) and considering normally-impinging light polarized in the direction indicated in the inset of panel (a).
\label{Fig1}}
\end{figure}

\section*{RESULTS AND DISCUSSION} 

{\bf Classical PWFs in different morphologies.} The optical response of graphene nanoislands is well-described in the electrostatic limit, as their plasmon resonance wavelengths typically appear in the infrared regime, where the light wavelength is much larger than the plasmon wavelength of the material \cite{paper228,paper235}. In previous studies, an eigenmode expansion method has been adopted to express the linear optical response of a graphene nanostructure in terms of its supported plasmon modes \cite{paper228,paper235}. Alternatively, one can associate a plasmon mode $j$ with its induced charge distribution, which we refer to as the PWF of mode $j$ \cite{paper257}. We demonstrate the power of the PWF formalism in Fig.\ \ref{Fig1}. In the spectral region dominated by the lowest-order plasmon mode ($j=1$) supported by graphene islands of varied morphology, we compare extinction spectra predicted in the analytical PWF description (solid curves, details in Methods) with those obtained upon fully-numerical finite-element solution of Maxwell's equations (dashed curves, COMSOL). We find excellent agreement among analytical and numerical results, regardless of the nanostructure characteristic size $D$, graphene Fermi energy $\EF$, or dielectric permittivity of the surrounding environment $\epsilon$ (see colored labels). Here and in what follows, we describe the graphene surface conductivity in the Drude approximation, adopting a phenomenological inelastic damping rate $\hbar\tau^{-1}=20\,$meV (i.e., $\tau\approx33\,$fs, see Methods), unless specified otherwise.

\begin{figure}
\begin{centering}
\includegraphics[width=1\textwidth]{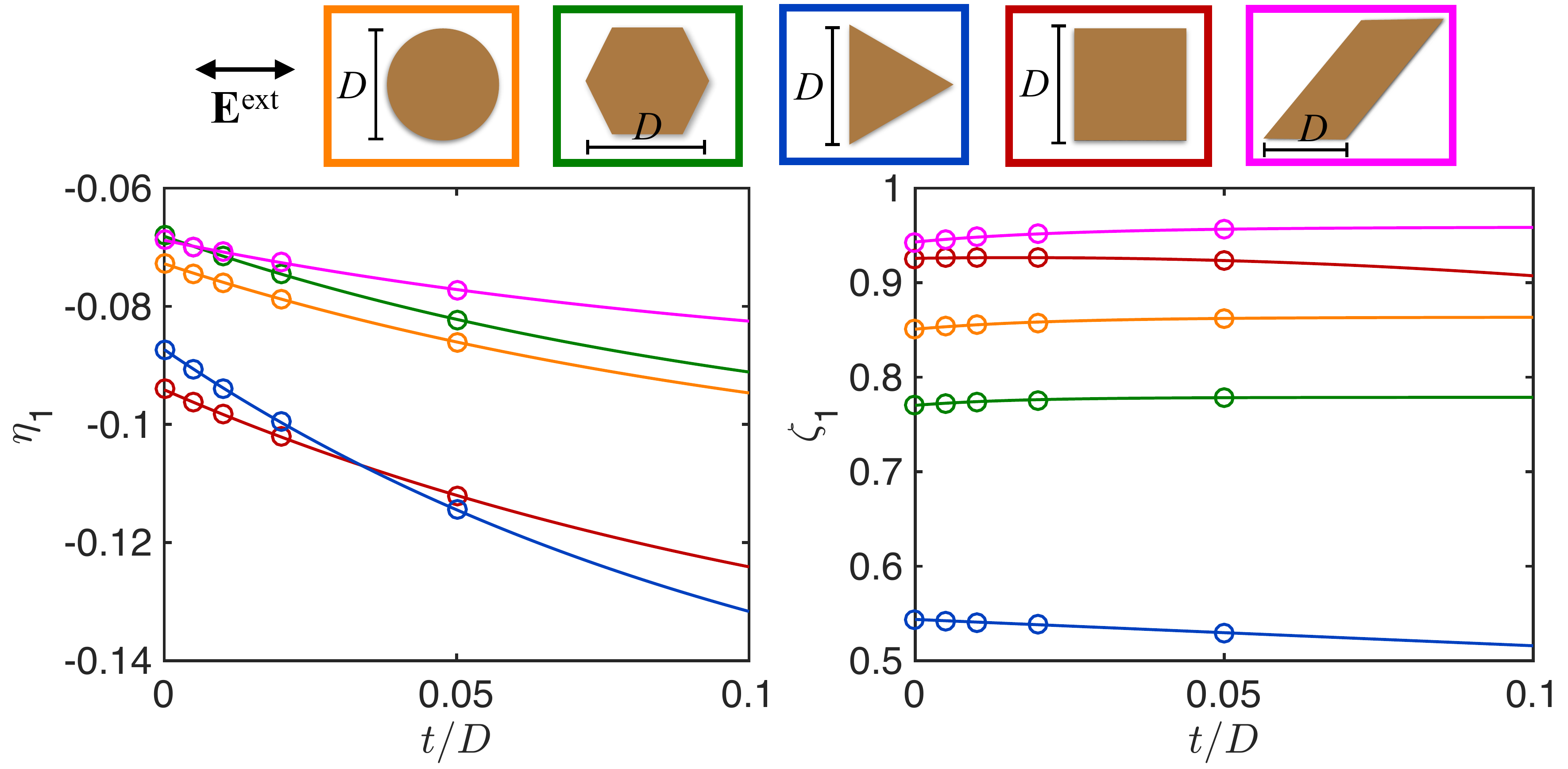}
\par\end{centering}
\caption{\textbf{Thickness dependence of the PWF analytical parameters.} We present the eigenvalue $\eta_1$ and normalized dipole moment $\zeta_1$ for the lowest-order dipolar plasmon mode of disks, triangles, ribbons, hexagons, and squares (see symbols, which are color-coded with the borders of the upper inset figures) as a function of the normalized effective graphene thickness $t/D$, where the characteristic size $D$ of a structure is indicated in the upper insets along with the light polarization direction (double arrow). Symbols for $\eta_1$ are obtained by fitting the numerically computed extinction spectra in the electrostatic limit, while $\zeta_1$ is calculated using Eq.\ (\ref{eq:zeta}). Solid curves, color-coded to the borders of the upper inset figures, correspond to the fitted expressions provided in Table\ \ref{Table1}.
\label{Fig2}}
\end{figure}

The analytical model used to produce the results presented in Fig.\ \ref{Fig1} is based on two parameters: the plasmon mode eigenvalue $\eta_j$ and dipole moment $\zeta_j$, where the index $j$ is a mode index and in this figure we focus on $j=1$, the lowest-order plasmon supported by each of the graphene islands under consideration. These two parameters are independent of the material properties, and in fact, they are determined by geometrical features alone. This means that the PWF treatment can also be used to describe other nanostructured 2D materials characterized by an isotropic surface conductivity. More precisely, the eigenvalue $\eta_j$ corresponds to the resonant value of the quantity $\eta=\ii\sigma(\omega)/D\omega\epsilon$, where $\sigma(\omega)$ is the graphene conductivity. Using the Drude model model for graphene conductivity (Eq.\ (\ref{Drude})), the plasmon frequency of mode $j$ can be analytically resolved in terms of the eigenvalue as\cite{paper235} $\omega\approx\omega_j-\ii\tau^{-1}/2$ with
\begin{align}
\omega_j=\frac{e/\hbar}{\sqrt{-\pi\eta_j\epsilon}}\;\sqrt{\frac{\EF}{D}}.
\label{wj}
\end{align}
Plasmons are then associated with negative eigenvalues $\eta_j<0$. This expression explains why the analytical model undergoes a minor redshift of the plasmon resonance peaks in all cases considered in Fig.\ \ref{Fig1} when the average surrounding permittivity $\epsilon$ increases from 1 to 1.35. We note that according to Eq.\ (\ref{wj}) the resonance positions for structures with the same value of $\sqrt{\EF/D\epsilon}$ should coincide, as illustrated by the green and blue curves.

The results of Fig.\ \ref{Fig1} correspond to islands of zero thickness. However, the charge that is optically induced on the graphene as the result of the excitation of a plasmon spans a finite thickness determined by the spatial extension of the out-of-plane carbon $p$ orbitals. A value of $\sim0.34\,$nm is typical used, corresponding to the inter-plane distance in graphite. Although this is an {\it ad hoc} parameter, it has been used in many prior studies of graphene plasmonics. Then, the thickness of the island $t$ has a finite value that can influence the plasmons. We thus present in Fig.\ \ref{Fig2} the dependence of the eigenvalue and dipole-moment parameters on the normalized thickness $t/D$. We find that the mode eigenvalue, which determines the plasmon resonance frequency, is more sensitive to variations of $t/D$, while the mode dipole moment $\zeta_1$ is relatively robust. We thus conclude that the spectral position of a plasmon resonance predicted for a graphene nanostructure described with a nonzero thickness is more prone to inaccuracy, unless a proper treatment of the thickness parameter is performed. Incidentally, the mode eigenvalues $\eta_1$ are obtained by fitting numerically-computed extinction spectra in the electrostatic limit for the lowest-order plasmon mode, while $\zeta_1$ is calculated using Eq.\ (\ref{eq:zeta}) (see Methods). We also provide $t/D$-dependent fits (solid curves in Fig.\ \ref{Fig2}) in Table\ \ref{Table1}. 

\begin{table}
\begin{centering}
\begin{tabular}{|c|c|c|c|c|c|}
\hline 
 & disk & hexagon & triangle & square & ribbon\tabularnewline
\hline 
$a_{\eta}$ & 0.03801 & 0.03846 & 0.07418 & 0.05537 & 0.02326\tabularnewline
\hline 
$b_{\eta}$ & -8.569 & -9.105 & -9.106 & -7.795 & -8.878\tabularnewline
\hline 
$c_{\eta}$ & -0.1108 & -0.1066 & -0.1615 & -0.1495 & -0.09208\tabularnewline
\hline 
$a_{\zeta}$ & -0.01267 & -0.008482 & 1 & -2.752 & -0.01572\tabularnewline
\hline 
$b_{\zeta}$ & -45.34 & -62.02 & -0.2826 & 0.09027 & -39.21\tabularnewline
\hline 
$c_{\zeta}$ & 0.8635 & 0.7787 & -0.4563 & 0.9258 & 0.9588\tabularnewline
\hline 
\end{tabular}
\par\end{centering}
\caption{Fitting functions for $\eta_{1}=a_{\eta}\exp\left(b_{\eta}x\right)+c_{\eta}$ and $\zeta_{1}=a_{\zeta}\exp\left(b_{\zeta}x\right)+c_{\zeta}$ corresponding to graphene islands of different morphologies as a function of the normalized thickness $x=t/D$. We apply these expressions for all morphologies considered in Fig.\ \ref{Fig2} except the graphene square, for which we have $\zeta_{1}=a_{\zeta}x^{2}+b_{\zeta}x+c_{\zeta}$.}
\label{Table1}
\end{table}

\begin{figure}
\begin{centering}
\includegraphics[width=1\textwidth]{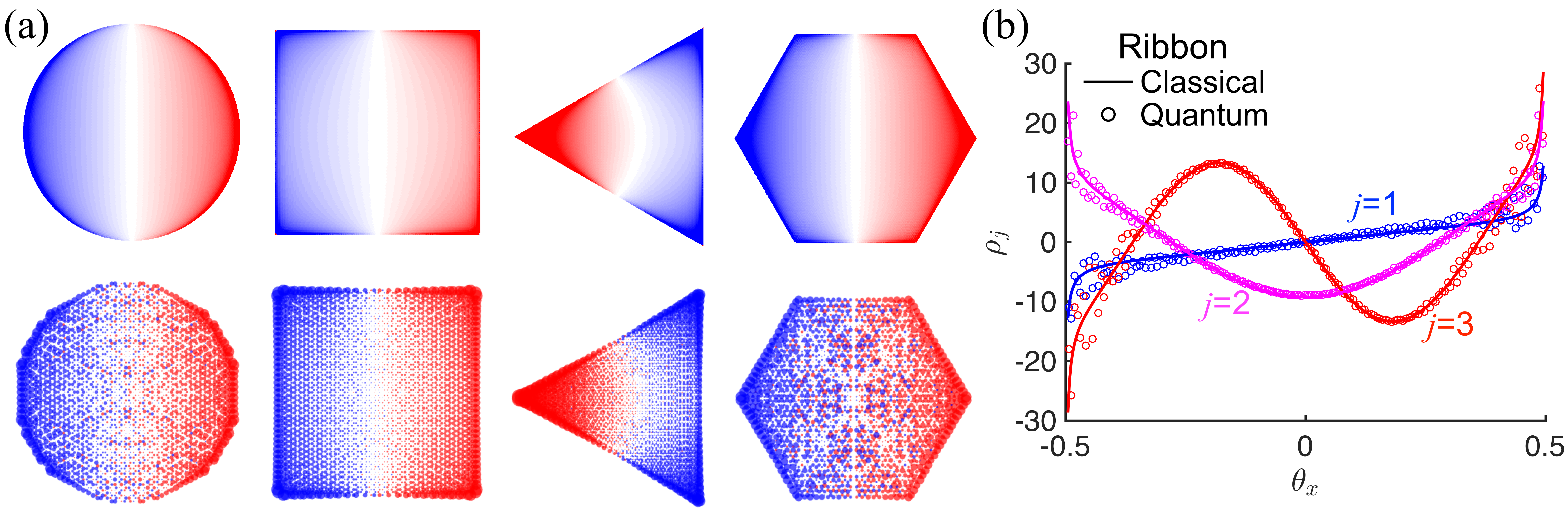}
\par\end{centering}
\caption{\textbf{Classical vs quantum PWFs.} (a) Density plots of the PWFs for the lowest-order ($j=1$) dipole mode are presented in the top row for several graphene geometries using the classical quastistatic model, whereas in the bottom row we show the induced charge distributions at the spectral position of the lowest-order plasmon resonance computed using an atomistic QM model (see Methods). Blue and red colors represent charges of opposite signs. (b) PWFs of the first three lowest-order ($j=1-3$, see labels) plasmon modes in graphene ribbons along the transversal ribbon direction \textit{x} are obtained using the classical model (solid curves). The induced charge distribution from the QM model is presented as symbols for the two dipole-active bright modes ($j=1,3$). The orientation of the incident light is along the ribbon width. In the quantum calculations we take $D=10\,$nm, $8.8\,$nm, $15\,$nm, $10\,$nm, and $20\,$nm for the disk, square, triangle, hexagon, and ribbon, respectively (see upper insets of Fig.\ \ref{Fig2}). These values correspond to $\sim2000-3000$ carbon atoms for the finite islands.
\label{Fig3}}
\end{figure}

{\bf Quantum-mechanical PWFs.} We present in Fig.\ \ref{Fig3} the spatial distributions of PWFs $\rho_1(\th)$ as a function of the normalized in-plane position vector $\th\equiv\Rb/D$ corresponding to the lowest-order plasmon mode contained in the collection of graphene structures considered here. In the upper row of Fig.\ \ref{Fig3}a, we present PWFs obtained using the classical model, with blue and red colors representing charges of opposite sign, so that the charge neutrality condition $\int d^{2}\th\,\rho_1(\th)=0$ is evident upon inspection. For comparison, the induced charge distributions of the same modes obtained from an atomistic QM model for graphene islands with lateral sizes on the order of $\sim10$\,nm are presented in the lower row of Fig.\ \ref{Fig3}a (see Methods for details on the QM model). We denote these induced charges associated with the plasmons as quantum PWFs (see Methods). The similarity between PWFs obtained from the classical and quantum models clearly indicates that the concept also holds in the quantum regime. In Fig.\ \ref{Fig3}b, PWFs for the first three lowest-order plasmon modes ($j=1-3$, see labels) are shown for a 1D graphene nanoribbon, where the $j=2$ mode, yielding $\zeta_2=0$, is a dark plasmon.

\begin{figure}
\begin{centering}
\includegraphics[width=0.7\textwidth]{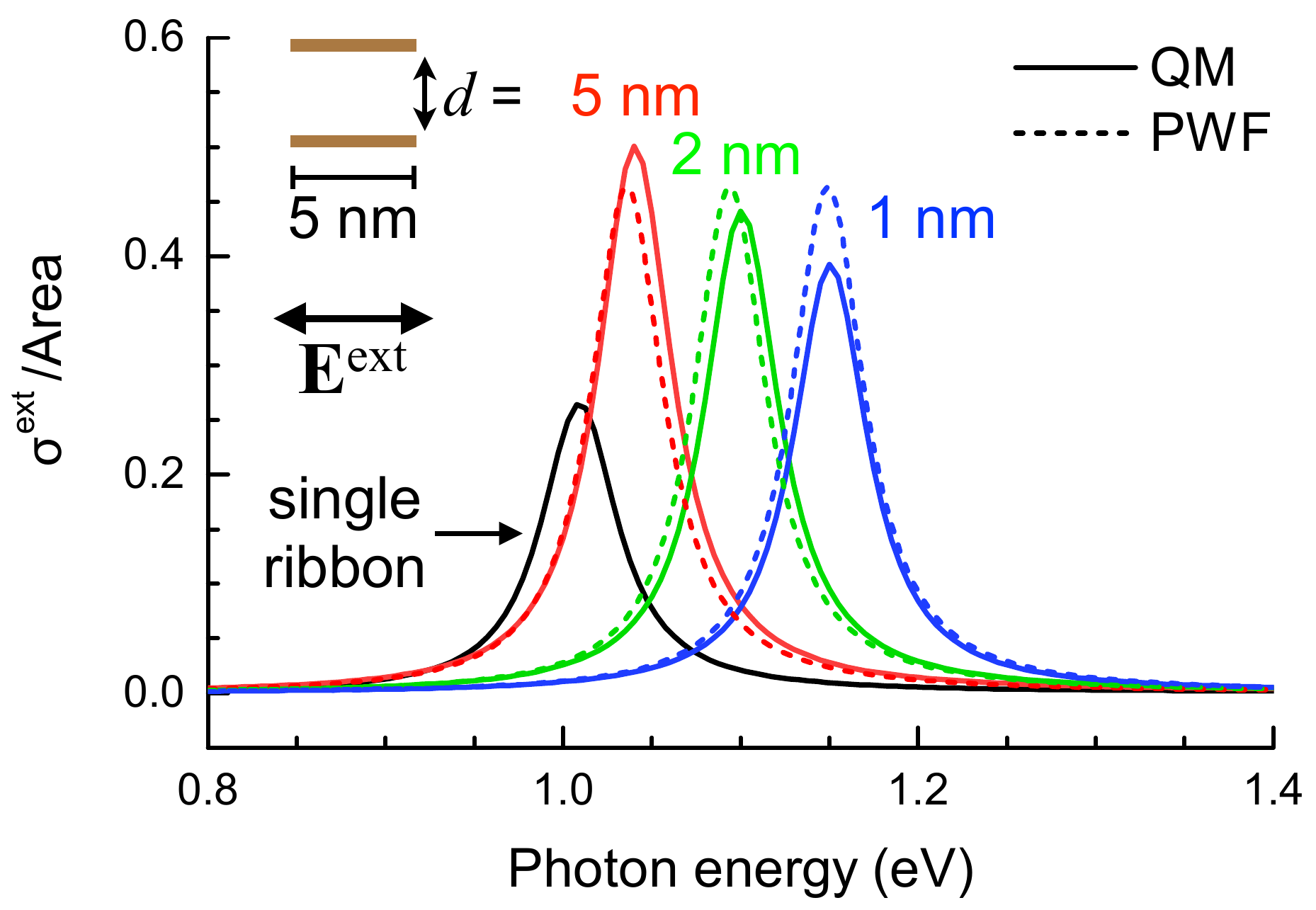}
\par\end{centering}
\caption{\textbf{Description of interacting nanoribbon plasmons using quantum-mechanical PWFs.} We compare the absorption cross-section of two vertically offset nanoribons (5\,nm width) for different separations $d$ (see color-coded labels) as described by direct QM simulations of the combined structure (solid curves) and using atomistic PWFs (dashed curves). Ribbons have armchair edges and their Fermi energy and damping energy are $\EF=1\,$eV and $\hbar\tau^{-1}=50\,$meV, respectively.
\label{Fig4}}
\end{figure}

Quantum PWFs are useful for studying the interaction between different graphene structures, avoiding costly numerical simulations. We put this concept to the test in Fig.\ \ref{Fig4}, where we present results for two parallel ribbons with a small vertical separation between them. In particular, we plot the extinction cross-section for transversal light polarization. The spectra are dominated by the lowest-order ribbon plasmon, which splits into two hybridized plasmons, one of which is dipole-active (i.e., it shows up in the spectra) and moves to the blue as the distance between ribbbons is decreased. The agreement between fully atomistic QM simulations (solid curves) and the PWF model (Eq.\ (\ref{rhonl}), broken curves), is rather satisfactory.

\begin{figure}
\begin{centering}
\includegraphics[width=0.4\textwidth]{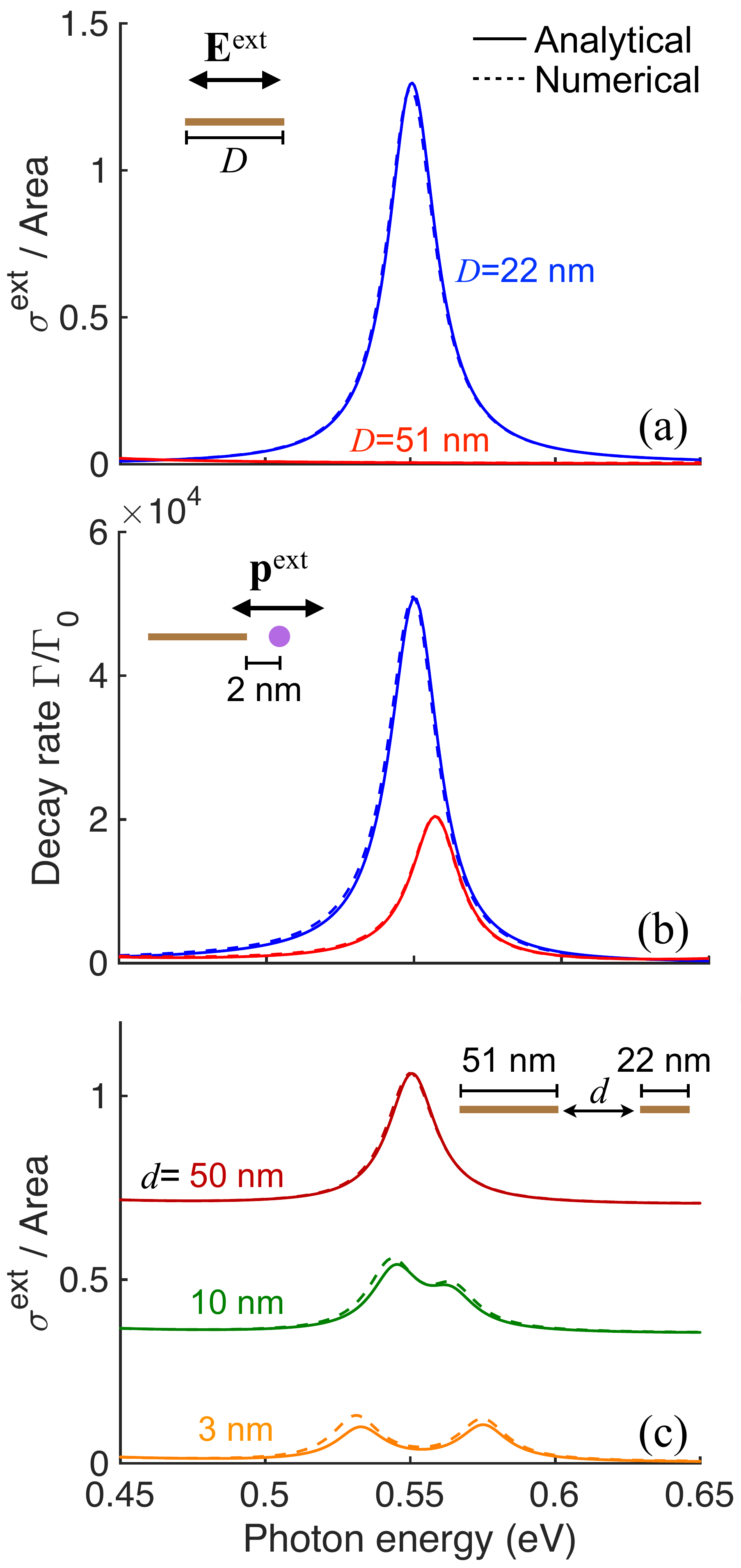}
\par\end{centering}
\caption{\textbf{Plasmon-induced transparency in paired graphene ribbons.} (a) Optical response of individual graphene ribbons of widths $D=22\,$nm (blue curves) and 51\,nm (red curves) under normal incidence with light polarization across the width of the ribbon. We assume $\EF=1\,$eV in all cases. The peak in extinction around 0.55\,eV corresponds to the lowest-order plasmon mode in the smaller ribbon, while the contribution from the larger ribbon is negligible in the frequency range shown. (b) Decay-rate enhancement for an external unit dipole $\pb^{\rm{ext}}$ placed 2\,nm away from the edge of the ribbons considered in panel (a). The dipole is oriented along the ribbon width. (c) Optical response of dimers composed of the two ribbons presented in panel (a) separated by an edge-to-edge distance $d$ in a co-planar configuration. When the separation distance $d$ decreases from 50 to 3\,nm (see labels), a transparency window appears around 0.55\,eV in the extinciton spectra as a result of the interaction between the $j=1$ mode (lowest-order dipole mode) in the smaller ribbon and the $j=2$ mode (first dark mode) in the larger one. Our analytical results (solid curves) agree well with numerical simulations (broken curves) in all cases.
\label{Fig5}}
\end{figure}

{\bf Plasmon-induced transparency.} As a way to demonstrate the versatility of the PWF formalism, we study the optical response of graphene structures interacting with external elements or with one another (see details in Methods). In Fig.\ \ref{Fig5}a we first present the optical extinction spectra of isolated graphene nanoribbons with different widths $D$, both of which are doped to $\EF=1$\,eV, and we include modes up to $j\leq3$ (see Fig.\ \ref{Fig3}b). In the frequency range shown, a prominent peak associated with the $j=1$ dipolar mode supported in the smaller ribbon ($D=22$\,nm, blue curves) appears around 0.55\,eV, whereas the contribution from the larger ribbon ($D=51$\,nm, red curves) is negligible at that energy. However, as shown in Fig.\ \ref{Fig5}b, where we simulate the decay rate of an oscillating unit dipole $\pb^{\rm{ext}}$ in the presence of either ribbon, a resonance feature appears in the spectrum for the larger one, which corresponds to its $j=2$ dark mode. This dark mode plays an important role when considering the optical response of a dimer formed by the co-planar combination of the two ribbons, with an edge-to-edge separation distance $d$, as shown in Fig.\ \ref{Fig5}c. The interaction between the bright and dark plasmonic modes in the small and large ribbons, respectively, results in a transparency window appearing around 0.55\,eV, which becomes more pronounced as the separation distance decreases from $d=50\,$nm to $d=3\,$nm. This phenomenon is known as plasmon-induced transparency, and has several applications, including slow light generation \cite{ZGW08,WCL14}. We note that results based on the PWF formalism are found to be in excellent agreement with fully-numerical simulations.

\begin{figure}
\begin{centering}
\includegraphics[width=1\textwidth]{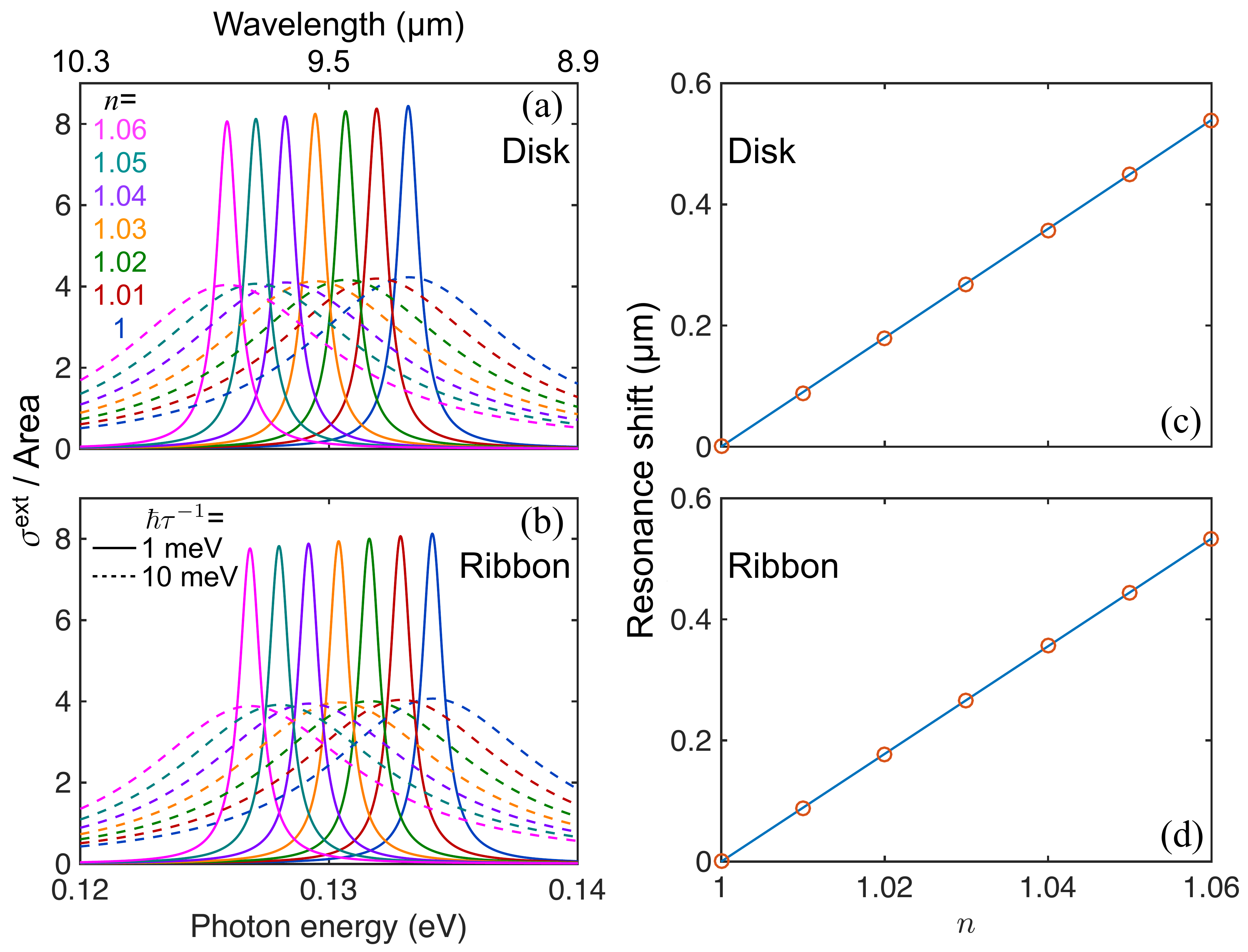}
\par\end{centering}
\caption{{\bf Refractive index sensing using individual graphene islands.} (a-b) Extinction spectra of a graphene disk ($D=120\,$nm, panel (a)) and ribbon ($D=125\,$nm, panel (b)) doped at $\EF=0.35\,$eV for different values of the refractive index $n=\sqrt{\epsilon}$ of the surrounding material (see labels). We present results for two different values of the inelastic damping rate ($\hbar\tau^{-1}=1\,$meV, $\tau\approx0.66\,$ps, solid curves; and $\hbar\tau^{-1}=10\,$meV, $\tau\approx66\,$fs, broken curves). (c-d) Dependence of the resonance shifts on the refractive index. Symbols correspond to the resonance positions extracted from the extinction spectra shown in panels (a-b). Solid curves present a linear fit, the slope of which is around $9\,\upmu$m/RIU for both cases. Here, we adopt local-RPA model for the graphene conductivity. The light incidence conditions conditions are the same as in Fig.\ \ref{Fig1}.\label{Fig6}}
\end{figure}

{\bf Sensing to the dielectric environment.} As another proof-of-concept demonstration, we apply the PWF formalism to simulate refractive index sensing assisted by mid-infrared plasmons in graphene nanostructures. In particular, we study in Figs.\ \ref{Fig6}a and \ref{Fig6}b the optical response of graphene disks ($\EF=0.35\,$eV, $D=120\,$nm) and ribbons ($\EF=0.35\,$eV, $D=125\,$nm), respectively, when they are immersed in media characterized by different values of refractive index $n=\sqrt \epsilon$. Here we adopt the local random-phase approximation (local-RPA) model for the graphene conductivity to properly account for the interband contribution to the optical response (see Methods), and we consider two different values of the phenomenological inelastic damping energy width, $\hbar\tau ^{-1}=1\,$meV and 10\,meV, corresponding to the results plotted as solid and dashed curves, respectively. In order to quantify the performance of the proposed sensor, we introduce the parameter $S\equiv\Delta\lambda /\Delta n$, which indicates the resonance wavelength shift per refractive index unit (RIU) change. From Fig.\ \ref{Fig6}c-d, we observe that both graphene disks and ribbons yield similar shifts of $\sim 9\,\upmu$m/RIU, independent of the chosen inelastic damping rate. The bulk figure of merit is defined as FoM$=S/$FWHM, where FWHM denotes the full width at half maximum of the resonance. We find a FWHM$\approx79\,$nm ($770$nm) for $\hbar\tau ^{-1}=1\,$meV (10\,meV) from Fig.\ \ref{Fig6}a-b, which gives a bulk FoM$\approx113.9$ (11.7). Note that the bulk FoMs calculated using either inelastic damping rate provide large values compared to others found in the literature \cite{WVK17,LYC15}. 

\section*{CONCLUSION}

In brief, this paper demonstrates the versatility of plasmon wave functions (PWFs) for studying the optical response of graphene structures with arbitrary morphologies. The present model is analytical and characterizes a plasmon resonance in a given geometry using only two real-valued parameters. The spatial distribution of PWFs calculated from classical modeling are found to be in excellent agreement with those obtained from atomistic quantum-mechanical simulations, even for structures of small ($\sim10$\,nm) lateral size when the edges are armchaired. We apply our analytical model to the study of graphene ribbon dimers, which accurately describes the plasmon-induced transparency that arises when bright and dark modes couple strongly. Additionally, the PWF formalism is used to explore graphene plasmon-assisted refractive index sensing at midinfrared frequencies, for which we predict a large bulk FoM (around 114), even when considering a conservative high inelastic damping rate. Finally, we note that the present analytical PWF formalism is universal and can be applied to model the optical response of other two-dimensional materials or thin films using their local 2D conductivitites as input.

\section*{METHODS}

\textbf{Graphene Conductivity.} In our classical approach, we characterize graphene through its optical surface conductivity, which is given in the local limit of the random-phase approximation (local-RPA) by \cite{GSC09,paper235}
\[
\sigma_{\mathrm{local-RPA}}\left(\omega\right)=\frac{e^{2}}{\pi\hbar^{2}}\frac{i}{\omega+i\tau^{-1}}\left[\mu^{T}-\int_{0}^{\infty}dE\frac{f\left(E\right)-f\left(-E\right)}{1-4E^{2}/\left[\hbar^{2}\left(\omega+i\tau^{-1}\right)^{2}\right]}\right].
\]
In the above expression $\mu^{T}=\mu+2k_{\mathrm{B}}T\log\left(1+\mathrm{e}^{-\mu/k_{\mathrm{B}}T}\right)$ is the thermally-corrected chemical potential, $\mu$ is the actual temperature-dependent chemical potential, $\tau$ is the inelastic relaxation time, and $f\left(E\right)=1/\left[1+\mathrm{e}^{\left(E-\mu\right)/k_{\mathrm{B}}T}\right]$. Analytical expressions for $\mu$ have been reported elsewhere \cite{paper287}, where it is shown that $\mu=\sqrt{\sqrt{\left(E_{\rm F}\right)^4+B^2\left(k_{\mathrm{B}} T\right)^4}-B\left(k_{\mathrm{B}} T\right)^2}$ with $B=\ln^2(16)/2\approx3.84$ constitutes a good analytical approximation that reduces to $\mu\approx\EF$ in the $\EF\gg k_{\mathrm{B}} T$ limit. Notice that from an experimental viewpoint, the doping conditions determine a temperature-independent carrier density $n$, which in turns controls the Fermi energy $\EF=\hbar\vF\sqrt{\pi n}$ (i.e., the chemical potential at zero temperature), where $\vF\approx10^6\,$m$/$s is the Fermi velocity in graphene. The local-RPA conductivity includes effects due to finite temperature, as well as both intra- and interband electron-hole-pair transitions in graphene. However, at room temperature and frequencies well-below $2\EF$, one can safely neglect temperature and interband effects, so that the expression above reduces to the Drude conductivity model,
\begin{align}
\sigma_{\mathrm{D}}\left(\omega\right)=\frac{e^{2}}{\pi\hbar^{2}}\frac{\ii\EF}{\omega+i\tau^{-1}},
\label{Drude}
\end{align}
Throughout this work, we adopt both Drude and local-RPA conductivities, assuming $T=300$\,K in cases where the latter is used.

\noindent \textbf{Classical Eigenmode Expansion and PWFs.} Following the formalism presented in Refs.\ \cite{paper228,paper257}, we intend to find the electric field $\Eb$ produced by a planar graphene structure in response to an impinging field $\mathbf{E}^{\mathrm{ext}}$, expressing it in frequency domain $\omega$ as the solution of the self-consistent equation
\begin{align}
\Eb\left(\Rb,\omega\right)=\mathbf{E}^{\mathrm{ext}}\left(\Rb,\omega\right)+\frac{i\sigma\left(\omega\right)}{\omega\epsilon\left(\omega\right)}\nabla_{\Rb}\int\frac{d^{2}\Rb'}{\left|\Rb-\Rb'\right|}\nabla_{\Rb'}\cdot f\left(\Rb'\right)\Eb\left(\Rb',\omega\right).\label{eq:EbR}
\end{align}
Here, $\epsilon$ is the average permittivity of the materials on either side of the graphene plane, while $f\left(\Rb\right)$ is a filling function that is 1 when the in-plane 2D position vector $\Rb$ lies within the graphene structure and 0 elsewhere (a vanishing positive number in practice). It should be noted that we are formulating the self-consistent electric field $\Eb$ in the graphene plane and the surface conductivity $\sigma$ can be computed using either Drude or local-RPA models. Defining the normalized 2D in-plane vectors $\th\equiv\Rb/D$ and $\fE\left(\th\right)\equiv D\sqrt{f\left(\th\right)}\mathbf{E}\left(\th,\omega\right)$, where $D$ is a characteristic length of the geometry under consideration (e.g., the side length of the graphene island), Eq.\ (\ref{eq:EbR}) can be recast as
\begin{align}
\fE\left(\th\right)=\fE^{\mathrm{ext}}\left(\th\right)+\eta\left(\omega\right)\int d^{2}\th'\;\Mb(\th,\th')\cdot\fE(\th'),\label{eq:eigeneq}
\end{align}
where $\eta\left(\omega\right)=i\sigma/(\omega D\epsilon)$ and
\[
\Mb(\th,\th')=\sqrt{f\left(\th\right)f\left(\th'\right)}\nabla_{\th}\otimes\nabla_{\th}\frac{1}{|\th-\th'|}
\]
is a real and symmetric operator. In consequence, $\Mb$ admits a set of real eigenmodes $\fE_{j}\left(\th\right)$ and eigenvalues $1/\eta_{j}$ defined through
\begin{align}
\fE_{j}(\th)=\eta_{j}\int d^{2}\th'\;\Mb(\th,\th')\cdot\fE_{j}(\th'),\label{eq:fieldexpansion}
\end{align}
such that the eigenmodes satisfy the orthogonality condition
\begin{align}
\int d^{2}\th\;\fE_{j}(\th)\cdot\fE_{j'}(\th)=\delta_{jj'}
\label{eq:ortho}
\end{align}
and the closure relation
\begin{align}
\sum_{j}\fE_{j}(\th)\otimes\fE_{j}(\th')=\delta(\th-\th')\mathbb{I}_{2},
\nonumber
\end{align}
where $\mathbb{I}_{2}$ denotes the $2\times2$ identity matrix in the sub-space of quasistatic electric-field solutions. Using the above eigenmodes, we write the solution to Eq.\ (\ref{eq:eigeneq}) as
\begin{align}
\fE(\th,\omega)=\sum_{j}\frac{C_{j}}{1-\eta\left(\omega\right)/\eta_{j}}\fE_{j}(\th),
\nonumber
\end{align}
where the expansion coefficients are given by
\begin{align}
C_{j}=\int d^{2}\th\;\fE_{j}(\th)\cdot\fE^{{\rm ext}}(\th,\omega).
\label{Cj}
\end{align}
From the closure relation, we have $\fE^{{\rm ext}}(\th,\omega)=\sum_{j}C_{j}\left(\omega\right)\fE_{j}(\th)$, which allows us to expresss the induced field as
\begin{align}
\fE^{\mathrm{ind}}(\th,\omega)=\sum_{j}\frac{C_{j}}{\eta_{j}/\eta\left(\omega\right)-1}\fE_{j}(\th).
\nonumber
\end{align}
We now define the PWF
\begin{align}
\rho_{j}(\th)\equiv\nabla_{\th}\cdot\sqrt{f(\th)}\fE_{j}(\th),
\label{PWFeq}
\end{align}
which corresponds to the induced charge distribution of the plasmon eigenmode $j$. Using the continuity equation along with Eq.\ (\ref{eq:EbR}), we can write the induced charge density $\rho^{\mathrm{ind}}$ as
\begin{align}
\rho^{{\rm ind}}(\th,\omega)=\frac{\epsilon}{D}\sum_{j}\frac{C_{j}}{1/\eta_{j}-1/\eta(\omega)}\rho_{j}(\th).\label{eq:indcharge}
\end{align}
Now, for a uniform electric field $\Eb^{\mathrm{ext}}$ associated with a light plane wave that acts on the graphene structure (we remind that $D$ is small compared with the light wavelength, so we can neglect the propagation phase in the incident field), we find, upon integration of Eq.\ (\ref{Cj}) by parts, $C_{j}=-\zetav_{j}\cdot\Eb^{\mathrm{ext}}$, where
\begin{align}
\zetav_{j}=\int d^{2}\th\rho_{j}(\th)\th\label{eq:zeta}
\end{align}
is a parameter that plays the role of the mode dipole moment. From the induced charge density, we calculate the induced dipole moment as
\[
\pb^{\mathrm{ind}}\left(\omega\right)=D^{3}\int d^{2}\th\;\rho^{\mathrm{ind}}(\th,\omega)\;\th,
\]
while comparing the above expression with the definition of the polarizability $\pb^{\mathrm{ind}}\left(\omega\right)=\alpha\left(\omega\right)\cdot\Eb^{\mathrm{ext}}$ and using Eq.\ (\ref{eq:indcharge}), we obtain the $2\times2$ in-plane polarizability tensor $\alpha\left(\omega\right)$,
\begin{align}
\alpha\left(\omega\right)=\epsilon D^{3}\sum_{j}\frac{\zetav_{j}\otimes\zetav_{j}}{1/\eta(\omega)-1/\eta_{j}}.
\label{alpha}
\end{align}
Finally, we calculate the extinction cross-section from the polarizability using the expression
\noindent 
\begin{align}
\sigma^{\mathrm{ext}}\left(\omega\right)=\frac{4\pi\omega}{c\sqrt{\epsilon}}\,\mathrm{Im}\left\{\alpha\right\}.
\nonumber
\end{align}
In summary, Eqs.\ (\ref{PWFeq}), (\ref{eq:zeta}), and (\ref{alpha}) allow us to calculate the far-field scattering properties of a graphene structure from the knowledge of its PWFs.

\noindent \textbf{Decay-Rate Enhancement.} The decay rate $\Gamma$ of a unit dipole $\pb^{\mathrm{ext}}$ oscillating at frequency $\omega$ and located at the position $\rb$ in an inhomogeneous space (e.g., in the presence of a graphene nanostructure) is given by \cite{NH06}
\begin{align}
\Gamma=\Gamma_{0}+\frac{2}{\hbar}\mathrm{Im}\left\{ \left(\pb^{\mathrm{ext}}\right)^{*}\cdot\Eb^{\mathrm{ind}}\right\},
\nonumber
\end{align}
where $\Gamma_{0}=4\omega^{3}\left|\pb^{\mathrm{ext}}\right|^{2}/3c^{3}\hbar$ is the dipole decay rate in free space. We evaluate this expression in the presence of a graphene island by integrating the induced charge (Eq.\ (\ref{eq:indcharge})) weighted by the Coulomb interaction to yield the induced electric field \[\Eb^{\mathrm{ind}}(\rb,\omega)=-\nabla_\rb\int d^2\Rb' \frac{\rho^{\rm ind}(\Rb'/D,\omega)}{|\rb-\Rb'|}\] evaluated at an arbitrary position $\rb$ from the PWF defined on the graphene island.

\noindent \textbf{Interaction between Islands.} We consider a system composed of multiple graphene structures, indexed by $\ell$ and centered at the positions $\rb_{\ell}$. We now define $\th\equiv(\rb_{\parallel}-\rb_{\ell})/D$, where $\rb_{\parallel}$ indicates the in-plane position vector of the corresponding island and $D$ is a characteristic normalization length. We also define the eigenvalue $\eta_{\ell j}$, eigenmode $\fE_{\ell j}$, PWF $\rho_{\ell j}$, and mode dipole moment $\zetav_{\ell j}$ for the plasmon mode $j$ associated with the corresponding graphene island $\ell$. Then, the self-consistent electric field $\fE$, having contributions from each island, can be expressed as $\fE=\sum_{\ell j}a_{\ell j}\fE_{\ell j}$. From Eqs.\ (\ref{eq:eigeneq})-(\ref{eq:ortho}), we obtain the self-consistent expression
\[
a_{\ell j}=\frac{1}{1-\eta_{\ell}(\omega)/\eta_{\ell j}}\left[C_{\ell j}+\eta_{\ell}(\omega)\sum_{\ell'\neq\ell}\sum_{j'}M_{\ell j,\ell'j'}a_{\ell'j'}\right]
\]
for the expansion coefficients $a_{\ell j}$, where $C_{\ell j}=-\zetav_{\ell j}\cdot\Eb^{\mathrm{ext}}$. Here,
\[
M_{\ell j,\ell'j'}=\int d^{2}\th\int d^{2}\th'\fE_{\ell j}(\th)\cdot\Mb(\th,\th')\cdot\fE_{\ell'j'}(\th')=-\int d^{2}\th\int d^{2}\th'\frac{\rho_{\ell j}(\th)\rho_{\ell'j'}(\th')}{|\th-\th'+\mathbf{d}_{\ell\ell'}/D|}
\]
describes the interaction between plasmon modes $j$ and $j'$ in two islands separated by a vector $\mathbf{d}_{\ell\ell'}=\rb_{\ell}-\rb_{\ell'}$. After solving for all $a_{\ell j}$'s, the total induced dipole moment can be expressed as
\[
\pb^{\mathrm{tot}}=\sum_{\ell j}\pb_{\ell j}=\sum_{\ell j}-\epsilon\eta_{\ell}D^{2}a_{\ell j}\zetav_{\ell j}.
\]
Eventually, the extinction cross-section of the whole system can be calculated as 
\begin{align}
\sigma^{\mathrm{ext}}\left(\omega\right)=\frac{4\pi\omega}{\sqrt{\epsilon}|\Eb^{{\rm ext}}|^{2}c}{\rm Im}\Big\{(\Eb^{{\rm ext}})^{*}\cdot\pb^{\mathrm{tot}}\Big\}.
\label{sigmaext}
\end{align}
Specifically, for the analytical results shown in Fig.\ \ref{Fig5}, we take modes $j=1-3$ (i.e., the PWFs displayed in Fig.\ \ref{Fig3}b) for each of the ribbons considered there. Incidentally, the integrals along the ribbon direction yields a logarithmic function times the infinite length of the ribbon, by which we divide the results in order to obtain a dipole per unit length.

\noindent \textbf{Atomistic QM Simulations and PWFs.} We adopt a previously established nearest-neighbor tight-binding model \cite{W1947,CGP09} to approximate the electronic structure of graphene islands. For a given structure, the resulting single-electron wave functions are inserted into the RPA susceptibility \cite{PN1966,HL1970} to calculate its optical response \cite{paper183}. To produce the results displayed in Fig.\ \ref{Fig3}, we have considered graphene hexagons, triangles, and ribbons that have exclusively armchair edge terminations, while the carbon sites in graphene disks and squares are arranged such that the geometry centers coincide with the center of a carbon-atom ring.

For a single graphene structure, the RPA description permits calculating the induced charge $\rho_l^{\rm ind}$ at each of its carbon atoms located at positions $\rb_l$. Assuming that the response to an external potential $\phi^{\rm ext}$ at optical frequency $\omega$ is dominated by the plasmons of the structure with frequencies $\omega_j$, linear response theory allows us to write the induced charge as \cite{PN1966}
\[
\rho_l^{\rm ind}=\frac{-e^2}{\hbar}\sum_{jl'}2\omega_j\frac{\rho_{jl}\rho_{jl'}}{\omega_j^2-\omega(\omega+\ii\tau^{-1})}\;\phi^{\rm ext}(\rb_l),
\]
where $\rho_{jl}$ is the transition charge density (from ground state to a one-plasmon state) associated with plasmon $j$ at the carbon atom $l$. For incident light with electric field $\Eb^{\rm ext}$, we have $\phi^{\rm ext}(\rb_l)=-\rb_l\cdot\Eb^{\rm ext}$. The set of numbers $\rho_{jl}$ play the role of a PWF, now described in an atomistic QM fashion. This concept is useful to account for the interaction between different graphene islands $n$, so that the effective external field experienced by each of them is the sum of the actual external field and the one produced by the rest of the islands. This idea leads to the expression
\begin{align}
\rho_{nl}^{\rm ind}=\frac{-e^2}{\hbar}\sum_{jl'}2\omega_{nj}\frac{\rho_{njl}\rho_{njl'}}{\omega_{nj}^2-\omega(\omega+\ii\tau^{-1})}\;\left[\phi^{\rm ext}(\rb_{l'})+\sum_{n'l''}v_{nl',n'l''}\rho_{n',l''}^{\rm ind}\right],
\label{rhonl}
\end{align}
where we have added labels $n$ to denote different graphene substructures and $v_{nl,n'l'}=1/|\rb_{nl}-\rb_{n'l'}|$ is the Coulomb interaction. For parallel ribbons under transversal polarization (Fig.\ \ref{Fig4}), $\rho_{nl}^{\rm ind}$ is repeated over all unit cells of each of the structures, so we need to sum  $v_{nl,n'l'}$ over cells and restrict $l$ and $l'$ to the first unit cell in Eq.\ (\ref{rhonl}). Because the total induced charge per cell is zero, we subtract the $1/r$ divergent part of the Coulomb interaction (independent of atom position relative the the unit cell center), which does not contribute to Eq.\ (\ref{rhonl}). From this equation, we obtain the total induced dipole $\pb^{\rm tot}=\sum_{nl}\rb_{nl}\rho_{nl}^{\rm ind}$, and from here the extinction cross-section using Eq.\ (\ref{sigmaext}).

\section*{Acknowledgments}

This work has been supported in part by the Spanish MINECO (MAT2014-59096-P and SEV2015-0522), the European Commission (Graphene Flagship 696656 and FP7-ICT-2013-613024-GRASP), Ag\`encia de Gesti\'o d'Ajuts Universitaris i de Recerca (AGAUR) (2014-SGR-1400), and Fundaci\'o Privada Cellex.


\end{document}